\documentclass[aps,prd,twocolumn,showpacs,superscriptaddress,nofootinbib]{revtex4-1}
\usepackage{bm}
\usepackage{amssymb,amsmath,amsthm}
\usepackage{hyperref}
\usepackage{mathrsfs} 
\usepackage[normalem]{ulem} 
\usepackage{graphicx}

\def\sech{{\mathrm{sech}}}

\newcommand{\eq}{\begin{equation}}
\newcommand{\eeq}{\end{equation}}
\newcommand{\be}{\begin{equation}}
\newcommand{\ee}{\end{equation}}
\newcommand{\bea}{\begin{eqnarray}}

\newcommand{\eea}{\end{eqnarray}}

\def\eref#1{(\ref{#1})}
\begin{document}
\title{Quantum vacuum radiation in optical glass}
\author{Stefano Liberati}
\affiliation{SISSA - International School for Advanced Studies, Via Bonomea 265, 34136, Trieste, Italy}
\affiliation{INFN, Sezione di Trieste}
\author{Angus Prain}
\affiliation{SISSA - International School for Advanced Studies, Via Bonomea 265, 34136, Trieste, Italy}
\author{Matt Visser}
\affiliation{School of Mathematics, Statistics, and Operations Research, Victoria University of Wellington, PO Box 600, Wellington 6140, New Zealand}
\begin{abstract}
 
A recent experimental claim of the detection of analogue Hawking radiation in an optical system [PRL {\bf105} (2010) 203901] has led to some controversy [PRL {\bf 107} (2011) 149401, 149402]. 
While this experiment strongly suggests some form of particle creation from the quantum vacuum (and hence it is {\em per se} very interesting), it is also true that it seems difficult to completely explain {all features of} the observations by adopting the perspective of a Hawking-like mechanism for the radiation.
For instance, the observed photons are emitted parallel to the optical horizon, and the relevant optical horizon is itself defined in an unusual manner by combining group and phase velocities.  This raises the question: Is this really Hawking radiation, or some other form of quantum vacuum radiation?  Naive estimates of the  {amount of quantum vacuum radiation generated due to the rapidly changing refractive index --- sometimes called the dynamical Casimir effect ---  are not encouraging.  However} we feel that naive estimates {could be} misleading {depending on the quantitative magnitude of two specific physical effects}: ``pulse steepening'' and ``pulse cresting''. 
{Plausible bounds on the maximum size of} these two effects results in estimates much closer to the experimental observations, and we argue that the dynamical Casimir effect is now {worth additional investigation}.

\end{abstract}                                           
\date{25 October 2011; \LaTeX-ed \today}
\maketitle

\section{Background}

It has been theoretically demonstrated that electromagnetic field modes propagating in a dielectric medium with a spacetime dependent refractive index can closely resemble field modes in a black hole background geometry~\cite{Philbin:2007ji}.  The idea was further developed (and analyzed in more realistic contexts) in the sequence of papers~\cite{Faccio:2009yw, Belgiorno:2010iz, Belgiorno:2010zz, Cacciatori:2010vr, Robertson:2011xp, Schutzhold:2011ze}.

The central idea of that work (and follow-up developments) is that a sufficiently intense, localized and moving refractive index perturbation (RIP) can give rise to a region in the interior of the perturbation where electromagnetic modes cannot propagate in the direction of motion of the perturbation.

That is, a mode slows down while climbing the trailing edge of the RIP, and if the RIP is sufficiently intense and fast, it is possible for the mode's group velocity to vanish in the frame of the RIP.  

The condition for the existence of such a ``blocking region'', associated with a RIP of peak refractive index $n_{\text{max}}=n_0+\eta$ traveling at velocity $v_{\text{RIP}}$ on top of a background refractive index $n_0$, would be 
\be
\frac{c}{n_{0}+\eta}<v_{\text{RIP}}<\frac{c}{n_0}. \label{E:condition1}
\ee
This can equivalently be rephrased as a rather tight constraint on the refractive index
\be
n_0 \in\left[\frac{c}{v_{\text{RIP}}}-\eta, \frac{c}{v_{\text{RIP}}}\right] . \label{E:condition1b}
\ee

 Recently, in Ref.~\cite{Belgiorno:2010wn}  {Belgiorno et al.}~reported an ingenious experimental realization inspired by this proposal, and it was extensively argued that the observed photon emission could be interpreted as the analogue of spontaneous quantum Hawking radiation. {Additional  experimental and theoretical details are available in Ref.~\cite{Rubino}, and some critical analysis and response is available in Refs.~\cite{Schutzhold:2010am, Belgiorno:2010hk}.} Specifically, in this experiment an ultrashort laser pulse was used to create a traveling RIP in a transparent dielectric medium (fused silica glass) through the so-called non-linear Kerr effect (see e.g.~\cite{Philbin:2007ji}). The authors of  Refs.~\cite{Belgiorno:2010wn, Rubino} report experimental evidence of unpolarized photon emission, for which they suggest the most compelling explanation is a Hawking-like radiation due to the presence of the above mentioned blocking region which simulates a black hole-white hole system.

In closing this section, let us stress that this experiment should be contextualized in the wider framework of analogue models of gravity~\cite{Barcelo:2005fc} and that several other experiments have been recently performed (or planned) in this context. To cite a few, 
the classical stimulated Hawking emission experiment of Weinfurtner et al.~\cite{Weinfurtner:2010nu}, the hydraulic jump white hole of Jannes et al.~\cite{Jannes:2010sa}, and the negative phase velocity experiment of Rousseaux et al.~\cite{Rousseaux:2007is}. There is also a significant quantity of theoretical work on these more general analogue spacetimes, and their implications regarding the Hawking effect~\cite{Unruh:1980cg, Visser:1993ub, Visser:1997ux, Visser:1997yu, Unruh:2004zk, Barcelo:2006uw}.

\smallskip
Overall, we consider {this RIP-based} experimental scenario to be a very promising avenue in the ongoing search for a direct observation of some analogue of quantum vacuum radiation in condensed matter systems, and we hope that the analysis here will shed some light in this direction.

\vfill

\section{Experimental Situation}

The most salient features of the experimental observation are the following:
The fused silica was illuminated with a  laser pulse of 1~ps timescale,  and the input energy was varied in the 100--1200 $\mu$J range.
Radiation from the filament was then collected at 90 degrees with respect to the laser pulse propagation axis. This arrangement was chosen in order to strongly suppress, or eliminate, known spurious effects (e.g. Cherenkov radiation, fluorescence, etc.)~\cite{Belgiorno:2010wn}. 
The observed emission spectra were reasonably well fitted by Gaussian profiles. These Gaussian fits showed an increasing peak wavelength of emission, and increasing overall flux, with increasing pulse intensity. Characteristic numbers for the observations were the following: the emission is centered around a peak wavelength of about $(850~\pm~25)$~nm (see Fig.~3 of \cite{Belgiorno:2010wn}) and has a bandwidth that appears to depend on the pulse intensity. While the determination of the actual size of such a bandwidth is fitting dependent, it is evident that the most of the photon count is concentrated in the range from 700 to 1000~nm.

The interpretation of the observed spectrum as an optical analogue of spontaneous quantum Hawking radiation is based mainly on the above mentioned window condition.
In fact, while the observed radiation is clearly non-thermal, one can nevertheless claim it to be related to the optical horizons (blocking regions) via a modification of Eq.~\eqref{E:condition1}. However, this blocking condition, which at first glance seems rather straightforward, in fact contains a number of subtleties once optical dispersion is introduced. Indeed in situations of dispersive propagation with $n_0=n_0(\lambda)$, where $\lambda$ is now the wavelength of the specific mode under consideration, the simple refractive index blocking condition of Eq.~\eqref{E:condition1} becomes a wavelength-dependent  ``window condition'' defining those modes for which a blocking region exists:
\be
n_0(\lambda)\in\left[\frac{c}{v_{\text{RIP}}}-\eta, \frac{c}{v_{\text{RIP}}}\right] .\label{E:windowp}
\ee
The window condition \eqref{E:windowp} will be satisfied by a certain set of wavelengths, a set which depends on which form for the dispersion relation one chooses as well as on the experimental parameters such as the character of the RIP itself.  However, not all dispersion relations will necessarily lead to windows --- the ``window region'' might (and often does) define an empty set. 

The relevant wavelength dependence of the refractive index for fused silica is adequately approximated by the Sellmeier relation
\be
n(\lambda)=\sqrt{1+\sum^{m}_{i=1}\frac{B_i\,\lambda^2}{\lambda^2-\lambda_i^2}}.
\label{Sell}
\ee
(Except in the immediate vicinity  of resonances where the unphysical poles are regulated in real physical systems by other effects such as absorption.)  In Eq.~\eqref{Sell} the $B_i$ and $\lambda_i$ are a set of coefficients determined by matching experimental data for the dispersion. When truncating the sum to three poles, the data is best fit with the values
\begin{align}
\lambda_1=64.25\text{nm},\quad \lambda_2=114.00\text{nm}, \quad \lambda_3=9938.24\text{nm}.
\end{align}
which fix the location of the poles in the optical dispersion. 

Typical values of the induced perturbation in the refractive index $\eta$ are of the order of $10^{-3}$,  and higher, increasing with the intensity of the RIP. Calculating the $v_{\text{RIP}}$ relevant for the experiment requires not only the knowledge of the refractive index at the typical RIP frequency, but also the characteristic Bessel beam factor, which was measured to be $\theta_B=6.8^\circ$~\cite{Rubino, private}. The factor enters the formula for the velocity as a function of  refractive index as $v_{\text{signal}}=c/(q\,n_{\text{signal}})$ where $q=\cos(\theta_B)$. 

\section{Physical interpretation}

In Refs.~\cite{Belgiorno:2010wn, Rubino} the window condition of Eq.~\eqref{E:windowp} was applied with the understanding that $v_{\text{RIP}}$ was the group velocity of the pulse, whereas the the $n_0(\lambda)$ was taken to be the refractive index for {\em phase} velocities of modes in the fused silica. With this interpretation the window was found to be approximately  $(870,940)$~nm. 
We stress that this result is very strongly sensitive to the precise numbers used for the relevant quantities in the window condition of Eq.~\eqref{E:windowp} due to the relative flatness of the function $n_0(\lambda)$ over the range of wavelengths of interest here. 
{(The Cauchy approximation provides insufficient accuracy for determining the window condition and it is essential to use the Sellmeier approximation for this particular purpose.)}
Such a window appears to be in good agreement with the observed spectra,  hence supporting the authors' claim of the Hawking-like nature of the observed radiation. Furthermore, the peak shift and spectral broadening with intensity are also claimed to be expected in this interpretative framework. In particular, the emission bandwidth is predicted to depend on $\eta$ which in turn is a linear function of the pulse intensity.

While these aspects of the evidence are surely supportive of a Hawking-like interpretation for the origin of the observed radiation, it was immediately noticed by several researchers in the analogue gravity community that certain other features of the experiment are problematic from this point of view (see e.g. the Comment~\cite{Schutzhold:2010am},  and Reply \cite{Belgiorno:2010hk}). 

We do not wish here to further enter into that on-going debate,  but we shall instead limit ourselves to stressing two points of main concern from our point of view:
\begin{itemize}
\item 
First of all we find odd the observation that photons are emitted at 90 degrees with respect to the ostensible horizons set by the window condition \eqref{E:condition1}. In fact, the window condition only holds for modes aligned with the RIP direction of propagation, so one would not expect at all in this model any photon production at 90 degrees to the beam axis. 
\item
Furthermore, we think that some concern has to be raised with the application of the window condition itself, as it rather oddly mixes the group velocity of the RIP with the \emph{phase} velocities of the vacuum modes in a dispersive medium.  On the contrary, we feel that there are good physics arguments for expecting the window condition to be determined by only considering group velocities. 
\end{itemize}

Specifically, let us consider an observer in the rest frame of the pulse as it moves down through the glass at velocity $v_\text{RIP}<c$. Then a probe signal chasing the RIP from behind will be seen to be approaching with a velocity given by the relativistic formula for the combination of velocities
\be
v_\text{probe}=\frac{ v_\text{RIP}\pm v_\text{probe} }{1\pm v_\text{RIP}v_\text{probe}/c^2  }.
\ee 
Including the effect of a Bessel beam factor, we have $v_\text{RIP}=c/(q n_\text{RIP})$. Then 
\be
\frac{v_\text{probe}}{c}=\frac{n_\text{probe}-qn_\text{RIP}}{qn_\text{RIP}n_\text{probe}-1}.
\ee
From this we see that, in the optical band, zero approach velocity (in the RIP frame) is achieved at the location $x^*$ such that
 \be
 n_\text{probe}(\lambda,x^* )=qn_\text{RIP}=qn(\lambda_\text{RIP}). \label{window_velocities}
 \ee
This defines a wavelength-dependent blocking region since the probe sees the presence of the RIP as a local increase in the ambient refractive index.  Since $q<1$, and these are interpreted as group refractive indices, only faster wave packets (ones with a smaller group refractive index, including the Bessel beam factor $q$) possess a horizon. Since $n_{\text{probe}}(\lambda,x)$ varies between the background value $n_0(\lambda)$ and the value at the peak of the RIP $n_0(\lambda)+\eta$, the above condition Eq.~\eqref{window_velocities} is equivalent to Eq.~\eqref{E:windowp}. 
 
{However, from a theoretical perspective, use of the relativistic combination of velocities formula seems to us to be well-motivated only if} $v_\text{RIP}$ and $v_\text{probe}$ are either both group {velocities} or both phase velocities. Certainly $v_\text{RIP}$ is a group velocity since the RIP is a soliton wave packet. This rather strongly suggests that in determining the window region $v_\text{probe}$ should also be interpreted as a group velocity. 

If we now adopt a group velocity interpretation of the window condition, we are obliged to consider the \emph{group velocity refractive index}  which is defined as
\be
n_g(\lambda):=\frac{d[\omega n(\omega)]}{d\omega}=n+\omega\frac{dn}{d\omega}=n-\lambda\frac{dn}{d\lambda},
\ee
where $\omega$ is a frequency. Using this formula the window condition becomes
\be
n_g(\lambda)\in\left[\frac{c\,q}{v_{\text{RIP}}}-\eta, \frac{c\,q}{v_{\text{RIP}}}\right].
\ee
The problem is that using this (more physically justified) formula, and the previously given values for the characteristic quantities, there is no range of wavelengths for which the window conditions is satisfied.  As a function of $\theta_B$ the window is non-empty only for $\theta_B\lesssim 3^\circ$ when $\eta=10^{-3}$ and $\lambda_\text{RIP}=1050$ nm. 

In summary,  we feel that these two facts alone are quite sufficient for motivating serious consideration of alternative physical explanations for the interesting observations reported in  Ref.~\cite{Belgiorno:2010wn}. Specifically, we shall argue here that the dynamical Casimir effect (particle production from the vacuum due to a time varying external field) is a competitive explanation that certainly deserves further theoretical and experimental investigation. 

In particular, we conjecture that the rapidly varying refractive index in the fused silica excites vacuum modes leading to substantial photon creation. This would be naturally isotropic in the subspace orthogonal to the direction of propagation of the RIP,  and would not be directly dependent on the above mentioned window condition.
In the following discussion we shall present an analytically solvable 1+1 model which, in spite of its simplicity, is able to reproduce many of the salient features of the experimentally observed photon emission.

\section{Setting up the calculation}

As is standard, we simplify the problem to the scalar electric field component of the full electromagnetic field, with the understanding that the results derived will apply to both photon polarizations independently, and will contribute half of the total flux expected from a full calculation.
The equation of motion for the $1+1$ dimensional electric field $\phi$ with a time variable refractive index $n$ is 
\be
\frac{1}{c^2}\left(n^2(t)\phi_{,t}\right)_{,t}=\phi_{,xx}.
\ee
In momentum space this reads
\be
\phi_{k, \tau \tau }+c^2k^2n^2(\tau)\phi_k=0, \label{E:origmode}
\ee
where we have introduced the ``conformal time" variable
\be
\tau(t)=\int^t\frac{dt'}{n^2(t')}.
\ee

For the refractive index we choose a $\sech^2(\cdot)$ time-dependent perturbation on top of momentum-dependent background value
\be
n^2(t)=n_0(k)^2+2\eta\; n_0(k)\;{\sech^2}(t/t_0)+\mathcal{O}(\eta^2),
\ee
where $\eta$ is a small parameter and $n_0(k)$ would be a model for the refractive index such as the Sellmeier approximation.  
This choice is motivated not only by exact solubility of the model, but also for its close approximation to the experimental Gaussian profile reported in \cite{Belgiorno:2010wn}.   

Then the equation of motion becomes
\be
\phi_{k,\tau\tau}+c^2k^2\left[n_0^2(k)+2\eta\; n_0(k)\;{\sech^2}(t(\tau)/t_0)\right]\phi_k=0, \label{E:modesech}
\ee
which can immediately be recognized as describing a time-independent Schroedinger scattering problem
\be
\psi_{,xx}+\frac{2m}{\hbar^2}\left[E-V(x)\right]\psi=0, \label{E:schro}
\ee
with 
\be
V(x)=V_0 \; \sech^2(x/x_0), \qquad m=\frac{1}{2}c^2k^2\hbar^2,
\ee
while
\be
E=n_0(k)^2, \qquad \text{and}\qquad V_0=-2\eta \;n_0(k).
\ee
This analogy allows us to directly write down the solutions to our equation of motion by appealing to well-known textbook results.

The Bogoliubov coefficient $\beta_k$ for our mode equation \eqref{E:modesech} is related to the transmission coefficient $T$ for the scattering problem \eqref{E:schro} by $|\beta_k|^2=1/T-1$. From the literature (see for example Ref.~\cite{Boonserm:2010px} and many references therein) we find the transmission probability
\be
T=\frac{\text{sinh}^2\left(\pi\sqrt{\frac{2mEx_0^2}{\hbar^2}}\right)}{{\sinh}^2\left(\pi\sqrt{\frac{2mEx_0^2}{\hbar^2}}\right)+{\cos}^2\left(\frac{\pi}{2}\sqrt{1-\frac{8mV_0x_0^2}{\hbar^2}}\right)},
\ee
so that
\be
|\beta_k|^2=\frac{{\cos}^2\left(\frac{\pi}{2}\sqrt{1+8\eta\Gamma_k^2/n_0(k)}\right)}{{\sinh}^2\left(\pi\Gamma_k \right)},
\label{E:betaexac}
\ee
with $\Gamma_k=ck\tau_0\,n_0(k)$.  The function $|\beta_k|$ rises from zero at $k=0$ to a maximum $\beta_{\text{max}}$ at $k_{\text{max}}$ before again decaying to zero for large $k$. Physically this structure reflects the transition from low $k$ modes which experience the RIP as a sudden, approximately instantaneous influence, to high $k$ modes for which the RIP represents an adiabatic process.  
There are also features at the poles (resonances) of the refractive index, but here we focus on the region between the resonances, which is the region of physical interest due to it being the only region where observations in the RIP experiments are preformed.  

Furthermore, as we will show below, all of the features of the function $|\beta_k|$ deriving from the time dependence of $n(t)$ occur for values of $k$ for which $n_0(k)$ is essentially constant, whence the effects of dispersion can be ignored for the purposes of this calculation. Accordingly, we henceforth set $n_0(k)=n_0\simeq 1.458$ for what follows. 

Due to the smallness of $\eta$ we can expand the cosine about $\pi/2$ giving 
\be
|\beta_k|^2\simeq \eta^2\frac{4\pi^2c^4k^4t_0^4n_0^2}{\text{sinh}^2 (\pi ckt_0n_0)}-\eta^3 \frac{16\pi^2c^6k^6t_0^6n_0^3}{ \text{sinh}^2(\pi ckt_0n_0)}+\mathcal{O}(\eta^4)\label{E:shiftsech}\, ,
\ee
where we see explicitly the quadratic dependence of the spectrum on $\eta$ at lowest order. 

\section{Spectrum and flux}

The $3+1$ dimensional photon flux density is given by
\be
4\pi k^2|\beta_k|^2=\frac{ dN}{d\text{Vol}\;dk}.
\ee
Special care should be taken when converting to wavelengths (which is relevant for the comparison with the literature). 
The peak wavelength of emission $\lambda_{\text{peak}}$ is not simply given by $\lambda(k_\text{peak})$ since $\lambda_{\text{peak}}$ is the maxima of the function which enters the integral over $\lambda$, including the Jacobian factor
\be
\frac{dN}{d\text{Vol}\; d\lambda}=\left. \left|\frac{dk}{d\lambda}\right|\frac{ dN}{d\text{Vol}\;dk}=\frac{2\pi}{\lambda^2}\frac{ dN}{d\text{Vol}\;dk}\right|_{k=2\pi/\lambda.}
\ee

From the lowest order term in Eq.~\eqref{E:shiftsech} we get
\begin{align}
\frac{d N}{d\text{Vol}}=\int_0^{\infty}  F(\lambda) d\lambda \, ,
\end{align}
where 
\be
F(\lambda)=\frac{8}{15625}\frac{1}{\lambda^8}\frac{\pi^9c^4t_0^4n_0^2}{{\sinh}^2(2\pi^2 ct_0n_0/\lambda)}.
 \label{E:spectral_function}
\ee
Noting that the function $1/(u^8 {\sinh}^2(1/u))$ has its maximum at a point extemely close to $u=1/4$ we obtain
\be
\lambda_{\text{peak}}\simeq\frac{1}{2}\pi^2 ct_0 n_0 \, .
\ee

Above we have simply bootstrapped our initially 1+1 dimensional results to 3+1 dimensions using the interpretation of the equation of motion as applying to the magnitude $k$ of the wave vector $\bf{k}$. This is common procedure in dynamical Casimir effect calculations due to the isotropic nature of the emitted radiation. We have in mind the idea that the detector which observes the emitted photons actually watches only a small element of the silica glass which, as the pulse passes through, appears to the detector to undergo a rapid time variation in its refractive index.   Within our simple model this bootstrapping can be effectively achieved by integrating our Bogoliubov coefficient $|\beta_k|^2$ over a three dimensional $\bf{k}$ space, as we have done above. 

Calculating the total flux we have
\begin{align}
\frac{dN}{d\text{Vol}}=4\pi\int dk\; k^2|\beta_k|^2
=\frac{8\pi^2}{21}\frac{\eta^2}{c^3t_0^3n_0^5}.
\end{align}
The emitting region in this 3+1 dimensional interpretation would approximately be the cylindrical region defined by the width of the laser beam and length of  the region over which the refractive index is rapidly varying. Observationally the radius of the laser beam is $W\simeq 10^{-5}$~m \cite{private}, and we can estimate the length of the active region as $ct_0n_0$, so the physical volume of the active region is $V=\pi x_0^2(ct_0n_0)$. The total number of emitted photons is estimated to be
\be
N=\frac{8\pi^3}{21}\frac{\eta^2W^2}{c^2t_0^2n_0^4}\, .
\ee

\section{Data fitting}
%
The parameters given explicitly in Ref.~\cite{Belgiorno:2010iz} are 
\be
n_0=1.458, \quad t_0=2.5\times 10^{-14} \; \text{s},  \quad\eta=10^{-3} \label{numbers},
\ee
where $t_0$ is calculated from the reported initial spatial size of the pulse at its production $x_0\simeq 10^{-5}$~m and the RIP velocity.   Note that, up to a small perturbation, $\tau(t)=t/n_0^2$ which we use inside the $\sech^2(\cdot)$ function to maintain the integrability of the equation of motion. This effectively changes the above initial value of $t_0$ to $t_0/n_0^2=1.2\times 10^{-14}$~s.   

In principle these factors are all we need to derive our estimate of the peak frequency and flux. However, it is well known that the RIP shape is not invariant along its propagation (see e.g.~Ref.~\cite{Philbin:2007ji}). Indeed, there are two additional effects which are {certainly} necessary to take into account when fitting the data. These are ``pulse steepening'' and ``pulse cresting" which we shall discuss below.

\subsection{Pulse steepening}

{\em Pulse steepening} is a non-linear optical effect, akin to wave shoaling in water waves, which can be responsible for altering the size of the spatial region over which the pulse  goes from its background value to its peak value and back by a factor of ten or more. This effect works oppositely on the trailing and leading edges of the RIP; while the trailing edge becomes steeper the leading edge typically becomes shallower.   

An important question is what, if any, physical mechanism is responsible for setting an ultimate limit to the pulse steepening.   This point is not completely clear in the extant literature, although in many experimental situations one adopts the rule of thumb that pulse steepening saturates at ``about twice the carrier frequency'' \cite{Philbin:2007ji, private}.

We believe, however, that it is reasonable to expect that the limiting mechanism should be more closely related to physical properties of the silica glass. Furthermore we wish here to explore which effects can provide absolute limits to the pulse steepening and hence to the peak frequency predicted by our model. In this respect there are two obvious candidates. One is the {\em plasma frequency} $\omega_p$ and the other is optical absorption, known to be particularly effective near poles in the dispersion relation (here characterized in terms of the Sellmeier relation).  

 The plasma frequency definitely provides an ultimate limit to pulse steepening,  as it is fundamentally the electrons in the optical glass which communicate the propagating electromagnetic wave, and which themselves interact on a time scale given by the plasma frequency.  
From the Sellmeier relation, and the definition of the plasma frequency as the coefficient of the leading order deviation as the refractive index goes to $1$,
\begin{equation}
n^2 \to 1 - \frac{\omega_p^2}{\omega^2} = 1 - \frac{\lambda^2}{\lambda_p^2}\, ,
\end{equation}
one obtains
\begin{equation}
\omega_p = \sqrt{ B_i \; \omega_i^2}; \qquad \lambda_p = \frac{1}{\sqrt{\sum_i B_i /\lambda_i^2}}\, .
\end{equation}
For the suprasil glass used in the experiment one gets
\be
\omega_p=2.6 \times 10^{16}~\text{Hz}, \quad \lambda_p=72.7~\text{nm}\, .
\ee
With the above values one gets a typical timescale for the steepening of about $2\pi/\omega_p\simeq 2.4 \times 10^{-16}$ s.   This would be the ultimate limit to pulse steepening and we would expect the maximum steepness to closely approach, if not exactly saturate, this bound. 

The function $\sech^2(t/t_0)$ varies between its minimum and its maximum over a time scale of $2.4\times10^{-16}$~s when the parameter $t_0^{}$ is $2\times 10^{-16}$~s.


The above estimate represents an absolute upper bound to the steepening. However, one can argue that before the plasma frequency can start to play any role, pulse steepening will be limited by dispersive effects and in particular by the pole of the Sellmeier relation at approximately 114 nm (see Eq.~\eref{Sell}). Such poles represent resonances inside the silica glass and their role in limiting the time variation or propagating signals is essentially identical to that of the plasma frequency.  Nonetheless, if one takes, as we shall do, twice the time scale set by the plasma frequency $t_0\simeq 4\times 10^{-16}~\text{s}$, the resultant steepness is just below that set by the Sellmeier pole scale, exactly in the right ballpark to model a limiting scale set by the pole's presence.  Specifically the propagating RIP varies between its maximum and minimum over a length scale of approximately 140 nm when  $t_0\simeq 4\times 10^{-16}~\text{s}$ while the Sellmeier pole sits at 114 nm.

Let us here anticipate that to match the data our model requires that the pulse steepening saturates near the above scale $t_0$ rather than, say, at twice the carrier frequency given set by the above-mentioned rule of thumb.  This is a relevant point as it might be crucial in dismissing or supporting the present proposal against others.

\subsection{Crest amplification}

Another important effect which we expect to occur in a propagating and steepening pulse is \emph{crest amplification}, whereby the maximum amplitude of the pulse increases as the RIP steepens. The cause of this effect can be understood by a conservation of energy argument. First,  the integral of the intensity is constant (and given by the energy of the RIP). Second, due to the non-linear Kerr effect the intensity is proportional to the perturbation in the refractive index $\delta n(x)$. Therefore
\be
\int \delta n(x) \;dx=\text{constant},
\ee
so that 
\be
\eta \int [\text{shape of pulse}](x) \; dx =\text{constant}.
\ee
That is, if the pulse changes shape in such a way as to make the area under its normalized curve lower, we can expect the maximum amplitude, controlled by the parameter $\eta$ to increase.  Note that the total flux of emitted quanta is (approximately) quadratically dependent on $\eta$. 

While in the current context we lack any detailed understanding of the size of this effect, we argue that one can \emph{easily} expect $\eta$ to increase up to one order of magnitude. We shall hence use $\eta\simeq 10^{-2}$ in what follows. 

In the closely related but distinct context of fibre-optic amplifiers, power amplifications of 20 dB, 30 dB, and 40 dB are not uncommon~\cite{30dB-1, 30dB-2, 40dB}. These are power amplifications of $10^2$, $10^3$, and $10^4$ respectively, corresponding to amplitude increases of 10, 32, and 100 respectively. So a cresting factor of up to 100 (corresponding to $\eta\sim 10^{-1}$) is not inconceivable. 

Returning to the current context, the pulse steepening we have argued for, from $t_0\sim 10^{-14}$~s to  $t_0\sim 10^{-16}$~s, will (provided most of the pulse is concentrated in a narrow spike with a long shallow leading edge) squeeze the pulse temporally by up to a factor of 100, thereby potentially increasing the amplitude by a factor of up to 100.  Again, this implies that a cresting factor of up to 100 (corresponding to $\eta\sim 10^{-1}$) is not inconceivable. Therefore, we argue that our choice of  $\eta\simeq 10^{-2}$ is rather conservative.  
{It is important to note that crest amplification mainly affects the total flux, at $\mathcal{O}(\eta^2)$, and that effects on the peak wavelength and width of the spectrum are relatively subdominant, at $\mathcal{O}(\eta^3)$. }

\subsection{Peak frequency and Flux matching}

We are now in shape for computing the salient features predicted by our model.
From the previous calculation of the peak frequency we get (using the least amount of pulse steepening, that is the largest value of $t_0$ in the above given interval)
\be
\lambda_{\text{peak}}=\frac{1}{2}\pi^2 ct_0 n_0 \simeq 858~\mbox{nm}\, .
\ee
This is in striking agreement with the observational data.

{Note again, however, that to get this nice matching for the peak frequency it is essential that the pulse steepening occur on a timescale and distance scale close to those set by the pole in the Sellmeier relation discussed above.  In fact we have been generous and chosen a time scale slightly longer than the Sellmeier resonance time scale. This has the side effect of avoiding the resonance at $114$ nm --- which is the first resonance one encounters in the Sellmeier approximation, (moving up in frequency from the carrier frequency as the pulse steepens)}. Furthermore if one adopts the numerical model discussed in~\cite{Rubino}, then Fig.~10 of that reference implies a timescale of $\approx 1$~fs at the trailing edge of the RIP; this is within a factor 2${1\over2}$ of our preferred value.
In contrast, if one simply uses some small multiple of the carrier frequency as an estimate of the pulse steepening, the fit is quite bad.  It is therefore important to develop a clearer understanding of the amount of pulse steepening to be expected in this particular experimental setup.

On the other hand the total flux becomes
\be
N=\frac{8\pi^3}{21}\frac{\eta^2W^2}{c^2t_0^2n_0^4} \simeq 0.6 \quad \text {photons}.
\ee
This estimate should be supplemented by a factor two to take into account the two photon polarizations and with an additional solid angle scaling $\Omega/(4\pi)$ to account for the finite size detector.  {In the experiment the detector was a 5~cm diameter lens placed 10~cm from the filament~\cite{private}. In terms of the half-angle subtended by the detector one has $\Omega = 2\pi(1-\cos\theta)$ whence $\Omega/4\pi \approx 0.016$. } 
Unfortunately we do not have an accurate account of the observed flux.

\begin{figure}
\begin{center}
\includegraphics[scale=0.4]{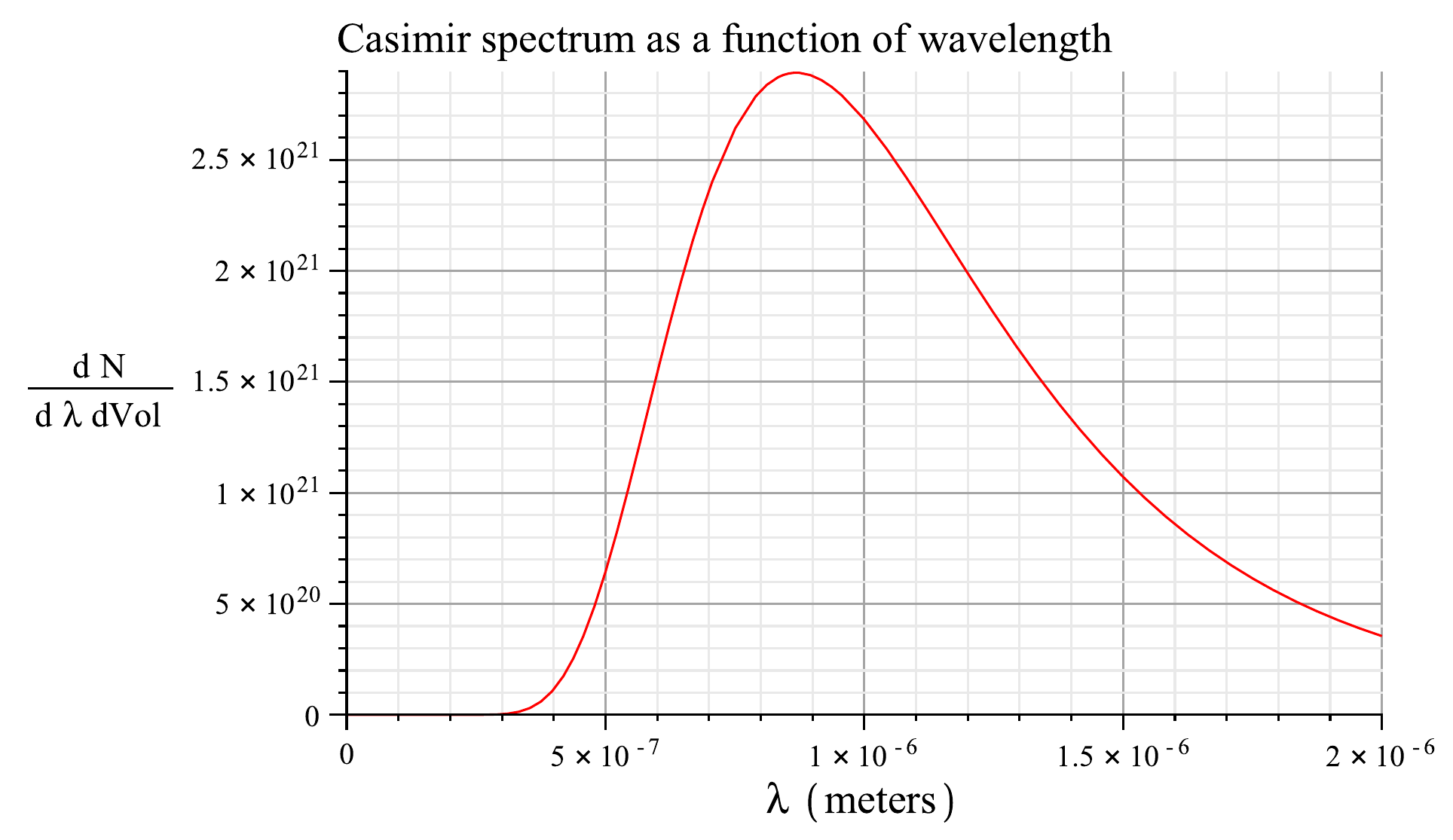}
\caption{Emission spectrum for the dynamical Casimir effect, as given by Eq.~\eqref{E:spectral_function} and using the parameter values discussed in the text: $n_0=1.45$, $t_0=4\times 10^{-16}$~s and $\eta=10^{-2}$. Recall that the total flux is given by the area under this curve multiplied by the volume of the emitting region. 
\label{F:spectrum_good_copy}}
\end{center}
\end{figure}

We consider the above numbers {to be somewhat} encouraging and supportive of the dynamical Casimir effect explanation. 
In Fig.~\ref{F:spectrum_good_copy} we show the emission spectrum for the dynamical Casimir effect as a function of wavelength as given by the function $F(\cdot)$ in Eq.~\eqref{E:spectral_function}. 

\subsection{Peak shift and broadening}

Of course, having shown that the relatively unsophisticated and crude model here proposed can can get anywhere near reproducing the observed peak frequency, and provide anywhere near a reasonable flux, while encouraging, is certainly not enough for claiming to have explained the experiment under investigation. In particular, we have not yet explained the abovementioned shift of the peak frequency and broadening of the spectrum. We shall now argue that our model can account for the former, and we shall critically discuss the relevance of the latter.

Experimentally, working from Fig.~3 in Ref.~\cite{Belgiorno:2010iz}, we can summarize the above two points with the following two empirical formulas:
\begin{equation}
\lambda_\mathrm{peak} \simeq 850\;\hbox{nm}+ 25\;\hbox{nm/mJ}\times E.
\end{equation}
\begin{equation}
\sigma_\lambda \simeq 100\;\hbox{nm/mJ}\times E.
\end{equation}
where $\sigma$ is the full-width-at-half-maximum (FWHM) of the Gaussian fits and $E$ is the input beam energy.

\subsubsection{Peak shift}

\begin{figure}
\begin{center}
\includegraphics[scale=0.4]{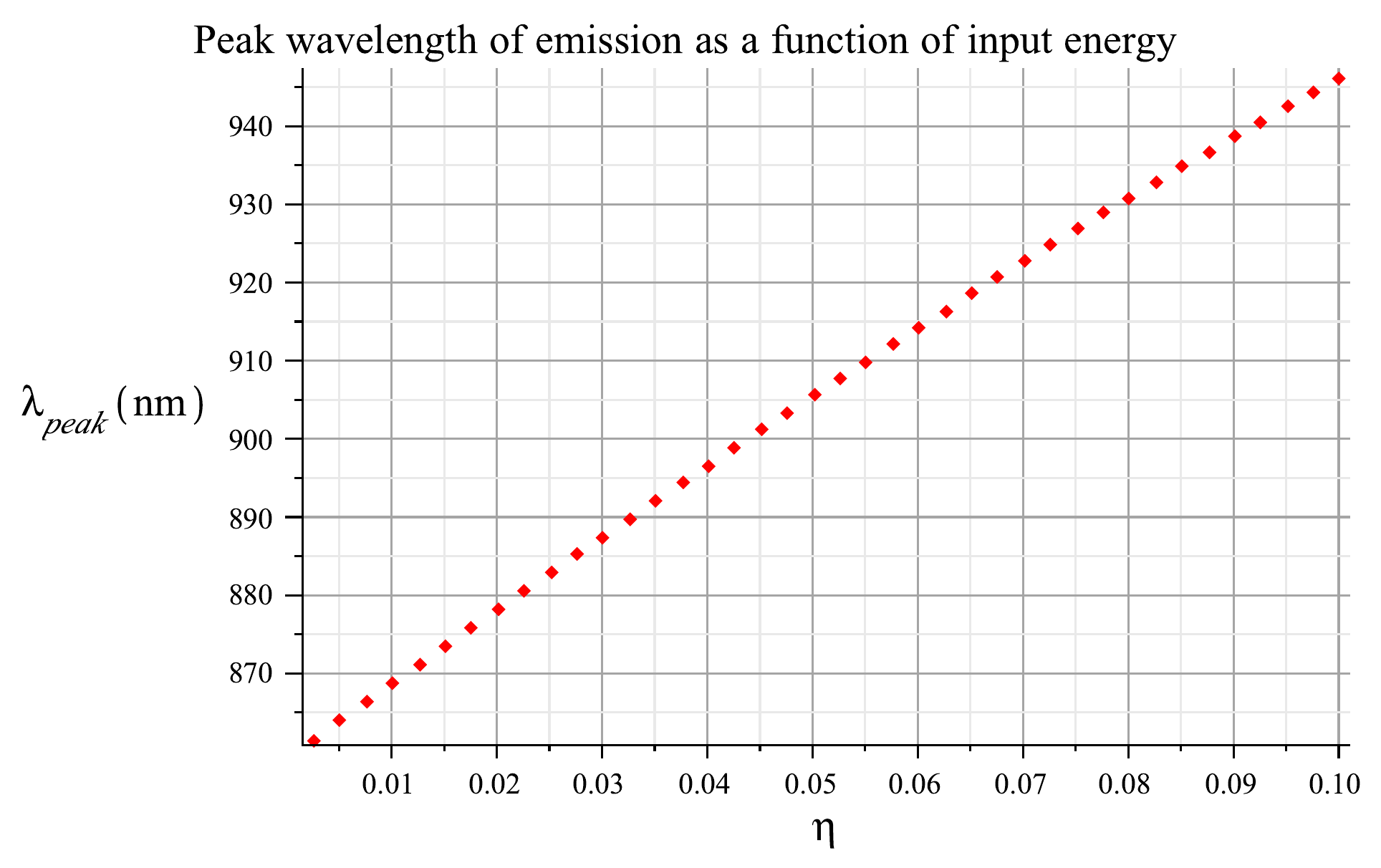}
\caption{Shift of the location of the maxima of the emission spectrum as a function of wavelength (in nm) for the sech$^2(\cdot)$ profile as the input energy is increased, here modeled by increasing the parameter $\eta$ from $2.5\times 10^{-3}$ to $10^{-1}$.
\label{F:shiftingsech}}
\end{center}
\end{figure}

From Eq.~\eqref{E:shiftsech} we see that the peak wavenumber (and hence also the peak wavelength of emission) depends on $\eta$ when we move to higher orders in the  $\eta$ expansion of the exact result.  In Fig.~\ref{F:shiftingsech} we plot the maxima of the spectral function $F(\lambda)$ for a range of values of $\eta$ between $0.25\times 10^{-2}$ and $10^{-1}$. 
Note that this shifting is consistent with the shift observed in the Belgiorno et al article~\cite{Belgiorno:2010iz}; increasing the input energy shifts the peak emission wavelength to a higher value.  We also note that the shifting is not exactly linear once we also take into account the $\eta^4$ term in the expansion of $|\beta_k|^2$. 

From Fig. 3 of  Ref.~\cite{Belgiorno:2010iz} one might argue that the peak wavelengths of emission are approximately
\begin{align}
&\lambda_{\text{peak}}\simeq 860 \hbox{ nm} \quad \text{at}\quad I=270\;\mu \hbox{J},\\
&\lambda_{\text{peak}}\simeq 875 \hbox{ nm} \quad \text{at} \quad I=1280\; \mu \hbox{J}.
\end{align}
That is, the experiments observe a 15~nm shift in peak wavelength from a factor of 5 increase of the input energy.  From our result shown in Fig.~\ref{F:shiftingsech} we see that our model predicts a spectral shift of about 10~nm from a factor of 10 increase of the input energy (given by $\eta$). 

We note that with an increase of the input energy by a factor of 10, our model predicts an increase in flux by a factor of (approximately) 100. In the case of a factor of 5 increase, as in the experimental data, we would expect from our model a flux increase of a factor of (approximately) 25. This predicted  behaviour is qualitatively consistent with the  observed spectra shown in Fig.~3 of  Ref.~\cite{Belgiorno:2010iz} where one sees an increase in peak flux from approximately 2 photoelectron counts at the lowest input energy (270 $\mu$J) to approximately 50 at the highest input energy (1280 $\mu$J).  

\subsubsection{Peak broadening}

\begin{figure}
\begin{center}
\includegraphics[scale=0.4]{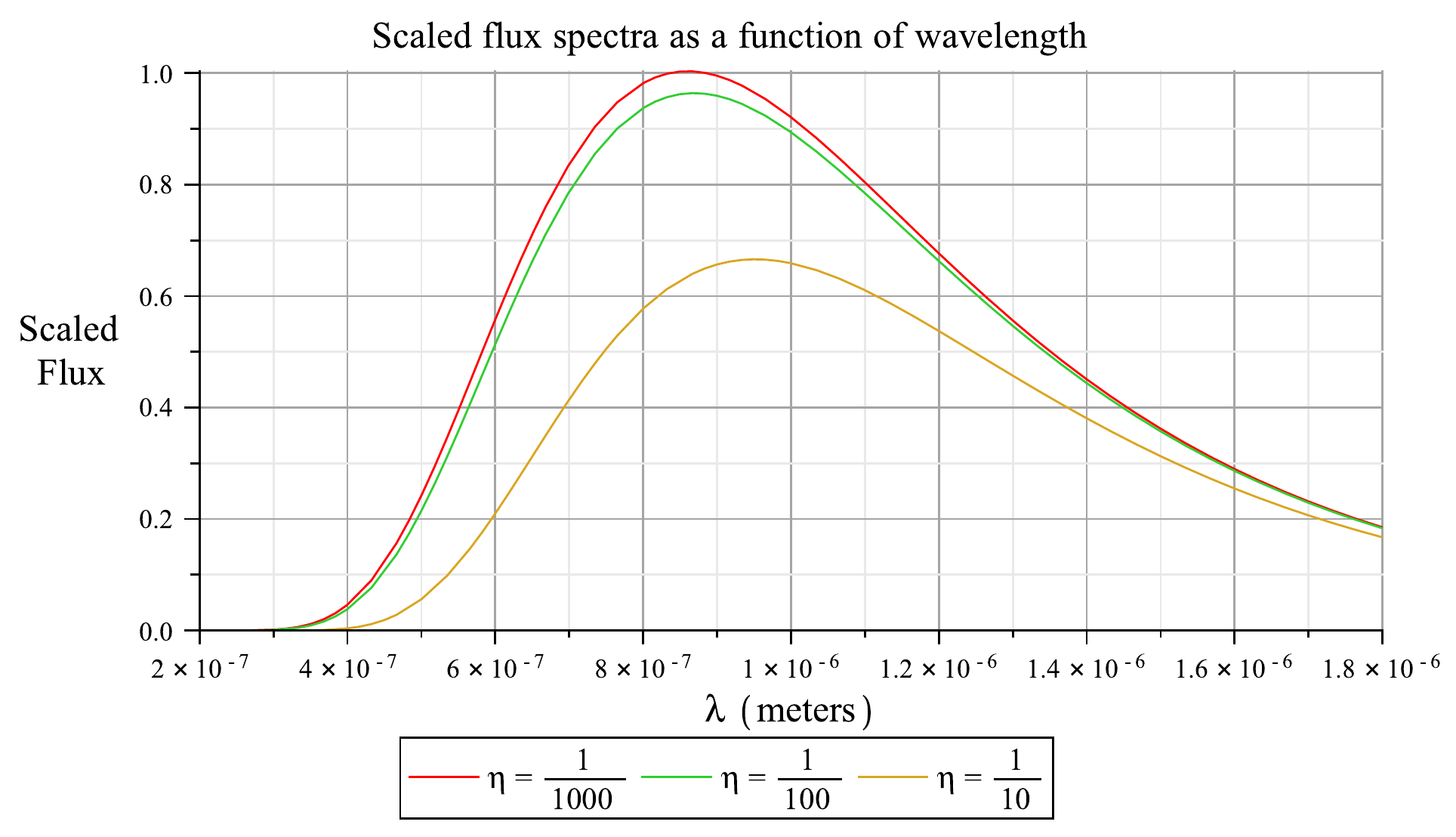}
\caption{Normalized spectra $F_\eta(\lambda)\times (\eta_0/\eta)^2 /[F_{\eta_0}(\lambda)]_\mathrm{peak}$.
Here $F(\cdot)$ is obtained from Eq.~\eqref{E:spectral_function}, but modified to include the higher order terms indicated in Eq.~\eqref{E:shiftsech}.  We normalize to the peak of the spectrum for $\eta_0=10^{-3}$, and display the result as a function of wavelength in meters. By multiplying by $(\eta_0/\eta)^2$, we can see how the shape and maximum of the spectrum as a function of $\eta$ is not exactly quadratic in $\eta$, but in fact depends on the higher order terms in indicated in \eqref{E:shiftsech}.    A numerical routine to find the FWHM bandwidths of these curves (which are independent of the amplitude) give the results shown in Eq.~\eqref{E:broadening_results}.  
\label{F:broadening}}
\end{center}
\end{figure}

Peak broadening, as we have mentioned above, is a feature observed in the data by matching the measured spectrum to Gaussian curves by eye. It is interesting to note that although the magnitude of the effect does not seem to be in agreement with the observational data, our oversimplified model also predicts a spectral broadening as the input energy ($\eta$) is increased.  We use  the FWHM definition for bandwidth, and find that $\sigma_\lambda$ behaves as
\begin{align}
&\sigma_\lambda\simeq 767\;\text{nm}\quad \text{for} \quad \eta=10^{-3}, \nonumber\\
&\sigma_\lambda\simeq 771\;\text{nm}\quad \text{for} \quad \eta=10^{-2}, \nonumber \\
&\sigma_\lambda\simeq 787\;\text{nm}\quad \text{for} \quad \eta=10^{-1}. 
\label{E:broadening_results}
\end{align}
As an increasing function of $\eta$, we see that the predicted broadening is in qualitative (though not quantitative) agreement with the data.  The broadening predicted by our simple model is shown in Fig.~\ref{F:broadening}, where we plot the scaled spectrum:
\begin{equation}
{F_\eta(\lambda)\over [F_{\eta_0}(\lambda)]_\mathrm{peak}} \times \left( {\eta_0\over\eta}\right)^2.
\end{equation}
That is, we plot the spectrum scaled to the maximum of the  spectrum at $\eta_0=10^{-3}$, multiplied by an extra factor of $(\eta_0/\eta)^2$. This is done in order for the spectral curves to be visually comparable, and to observe the non-quadratic dependence of the spectrum on $\eta$. Note that in this figure we have used an exaggerated range of values for the parameter $\eta$ to highlight the effect.  We also note that this figure shows the shifting peak wavelength of emission, where we see the peak moving to higher wavelengths as $\eta$ is increased. 
We again emphasize that the reported broadening in Fig.~3 of Ref.~\cite{Belgiorno:2010iz} arises from matching noisy data to purely phenomenologically chosen Gaussian curves (there is no physics reason underlying the choice of Gaussian curves), and that the reported magnitudes might, in a slightly more conservative analysis, be lower than those reported. 

\section{Conclusions}

 The model provided in this article is very simple, in many ways overly simple, but nevertheless provides a tolerable fit to many of the known aspects of the experimental situation. The dynamical Casimir effect envisaged in this model has a long and convoluted history in its own right (see for instance~\cite{Liberati:1998tf, Liberati:1998gs, Liberati:1998wg, Liberati:1999jq, Liberati:1999uw, Belgiorno:1999ha, Liberati:2000iq}). Particularly attractive features of the dynamical Casimir model are that it does not need any window, that there is no preferred direction, and that in performing the experiment one could look from any angle. 

 In the present context, the most pressing aspect of the model is to more fully understand pulse broadening, and to undertake a detailed confrontation of the model with empirical reality. {(Specifically, it would be important to develop precise quantitative estimates of the amount of pulse steepening and crest amplification to be expected in the relevant experimental context.)} Another potential line of development would be to go to a full 3+1 dimensional calculation --- we have already undertaken a perturbative 3+1 dimensional calculation but it taught  us nothing new beyond the present simple model. 

From a theoretical perspective, one might argue that the natural frame of reference for understanding this experimental result is the RIP frame.
Of course, in this frame the simple picture of a time dependent refractive index seems to be lacking, but one instead has to view the photons as being created by the optical glass rapidly flowing through the RIP; with the quantum ground state of the electromagnetic field being defined by the rest frame of the suprasil optical glass, quantum excitations are now generated by motion through an inhomogeneous medium. 

Furthermore, in the RIP frame, modes endowed with modified dispersion relation will see a sequence of subluminal regions of different intensity. 
Situations like this have been shown to still lead to particle production even in absence of any horizon/blocking region~\cite{Antonin:2011in, Finazzi:2011jd, Finazzi:2010yq, Finazzi:2010nc} for the ``dual" situation of modes subject to a supersonic dispersion relation in a ``flow" with two supersonic regions of different speeds. Indeed there seems to be mathematical correspondence that leads to the same behavior when passing from super- to sub-sonic in the dispersion relation and in the character of the flow regions~\cite{Finazzi-PhD-thesis}. If confirmed this analysis would corroborate our claim that a particle production from the quantum vacuum could still take place in our system even in the absence of analogue horizons.

 Finally, we emphasize that the dynamical Casimir effect is still quantum vacuum radiation, just not Hawking radiation.
Both the dynamical Casimir effect and Hawking radiation produce squeezed state outputs.
Direct tests of the difference between these two effects should focus on the different correlation structure in the emitted photons.  Specifically, it would be interesting to consider 2-photon correlations, and back-to-back photons (see  e.g.~\cite{Belgiorno:1999ha, Balbinot:2007de}).  These effects are likely to be different in detail.  One might also need to do considerably more than just looking at the squeezed state properties.  At a very practical level, the output radiation might not be ``clean'', due for instance to interaction with silica defects.  While much theoretical work remains to be done, in many ways the ball is now in the experimentalists court. 

\bigskip
\section*{Acknowledgments}
The authors wish to thank Francesco Belgiorno and Daniele Faccio for sharing information about their work, and for many useful comments and suggestions.
AP and SL wish to thank Stefano Finazzi for illuminating discussions and Silke Weinfurtner for useful remarks.
MV was supported by the Marsden Fund, administered by the Royal Society of New Zealand. 


\end{document}